\documentclass{emulateapj}

\usepackage{epsfig}

\begin{document}

\title{Constraints on the Size of Extra Dimensions from the\\
Orbital Evolution of Black-Hole X-Ray Binaries}

\author{Tim Johannsen}
\affil{Physics Department, University of Arizona, 1118 E. 4th Street, Tucson,
 AZ 85721}
\email{timj@physics.arizona.edu}

\author{Dimitrios Psaltis}
\affil{Physics and Astronomy Departments, University of Arizona,\\ 1118 E.
 4th Street, Tucson, AZ 85721}
\email{dpsaltis@physics.arizona.edu}

\author{Jeffrey E. McClintock}
\affil{Harvard-Smithsonian Center for Astrophysics, 60 Garden Street, 
Cambridge, MA 02138}
\email{jem@head.cfa.harvard.edu}

\begin{abstract}
One of the plausible unification schemes in physics considers the observable universe to be a 4-dimensional surface (the ``brane") embedded in a higher-dimensional curved spacetime (the ``bulk"). In such braneworld gravity models with infinitely large extra dimensions, black holes evaporate fast through the emission of the additional gravitational degrees of freedom, resulting in lifetimes of stellar-mass black holes that are significantly smaller than the Hubble time. We show that the predicted evaporation rate leads to a change in the orbital period of X-ray binaries harboring black holes that is observable with current instruments. We obtain an upper limit on the rate of change of the orbital period of the binary A0620$-$00 and use it to constrain the asymptotic curvature radius of the extra dimension to a value comparable to the one obtained by table-top experiments. Furthermore we argue that any measurement of a period increase for low-mass X-ray binaries with a high mass ratio is evidence for new physics beyond general relativity and the standard model.
\end{abstract}


\section{Introduction}

In the search for the unified theory of all forces, an essential ingredient is the solution of the so-called hierarchy problem. The fundamental scale of gravity, the Planck mass, exceeds the electroweak scale by 16 orders of magnitude. In order to resolve this discrepancy, Arkani-Hamed, Dimopoulos, \& Dvali (1998) suggested that gravity is allowed to propagate in more then three spatial dimensions and is hence ``diluted'' in our universe. This leads to modifications of gravity at distances that are smaller than those probed by experiments. Indeed, Newton's inverse square law has been tested down to the sub-mm range (Kapner et al. 2007; Geraci et al. 2008), hence verifying that our space is three-dimensional at macroscopic scales. Any modification of gravity involving extra dimensions therefore has to ensure that additional space dimensions only effect our world at distances that are smaller than those experimental limits.

Braneworld gravity offers a solution to this problem in the form of two different scenarios. One approach (Arkani-Hamed et al. 1998) is to compactify $n$ extra dimensions at scales smaller than those set by experiment. The fundamental Planck mass can be pushed down to the electroweak scale of about 1 TeV, provided the extra dimensions are large enough. For $n$ extra dimensions, the limit is $R\lesssim 10^{30/n-17}$ cm. For $n\geq2$, extra dimensions would have a sub-mm size, which is just at the limit up to which the inverse square law has been verified. This model can also be embedded in string theory (Antoniadis et al. 1998). However, it cannot be tested in astrophysics, because those length scales are well below astronomical distances.

A second scenario (Randall \& Sundrum 1999) is based on a different idea. The four-dimensional brane with all standard model particles is embedded in an infinite five-dimensional anti-de Sitter space. Deviations from the inverse square law, however, only manifest at distances smaller than the asymptotic curvature radius $L$ of the bulk because the latter is filled with a negative cosmological constant. This setup has dramatic implications for astrophysical black holes.

No stable solutions for black holes on the brane have been found to date. Numerical integration of the classical bulk equations governing the evolution of black holes in the RS2 scenario that are localized on the brane indicated that black holes are unstable and hence lose energy in the extra dimension (Tanaka 2003). Based on the AdS/CFT correspondence, Tanaka (2003) suggested that stable black holes may not exist on the brane at all. Applying the AdS/CFT correspondence to AdS braneworld  models, Emparan, Fabbri, \& Kaloper (2002) conjectured that black holes localized on the brane that are solutions of the classical bulk equations in $AdS_{\rm D+1}$ with the brane boundary conditions correspond to quantum-corrected black holes in $D$ dimensions. Black holes can then evaporate through the emission of a large number of CFT modes with a lifetime given by (Emparan, Garc\'{i}a-Bellido, \& Kaloper 2003; see, however, Fitzpatrick, Randall, \& Wiseman 2006)\begin{equation}\tau\sim1.2\times10^2\left(\frac{M}{M_{\odot}}\right)^3\left(\frac{1~\rm{mm}}{L}\right)^2 \rm{yr},\end{equation} which is only of the order of a hundred thousand years for black holes with a mass $M$ of a few solar masses and an asymptotic curvature $L$ in the sub-mm range. Therefore, astrophysical black holes can radiate away most of their mass at cosmologically relevant timescales. This property has been used to constrain $L$ from a kinematic limit on the age of the black hole XTE J1118+480 (Psaltis 2007a), yielding $L<80~\rm{\mu m}$, as well as to give a possible explanation of the unusual observed black-hole mass function (Postnov \& Cherepashchuk 2003).

In the Randall-Sundrum model RS2, the gravitational potential at distances close to $L$ takes the form (Randall \& Sundrum 1999)\begin{equation}V(r)\approx-G\frac{m_1 m_2}{r}\left(1+\frac{L^2}{r^2}\right).\end{equation} Adelberger et al. (2007) report a 1$\sigma$-upper limit on $L$ of $11~\mu\rm{m}$. A 3$\sigma$-constraint has not been computed yet, but it should be significantly larger and comparable to the 95\%-confidence upper bound of $44~\mu\rm{m}$ for the size of one compact extra dimension (Kapner et al. 2007). Hereafter, we will take the latter as the current experimental constraint on $L$.

In this paper, we constrain the asymptotic curvature radius $L$ in the RS2 scenario by considering the evaporation of black holes in X-ray binaries. A mass loss of the black hole in the extra dimension leads to an evolution of the orbit, which is potentially measurable. Competing effects are the orbital period evolution caused by magnetic braking and the evolution of the companion star (see, e.g., Verbunt 1993). In \S 2, we systematically derive the rate of change of the orbital period involving all three effects as well as non-conservative mass transfer. We present the results in \S 3, where we identify systems with predominant black-hole evaporation. In \S 4 we focus on the black-hole binary A0620$-$00 in particular and obtain an upper limit of $L<161~\rm{\mu m}$, which is already comparable to the limit from table-top experiments. In the final section (\S 5) we discuss the potential of this binary as well as of other sources to constrain $L$ down to a few microns.

\section{Orbital Evolution of a Black-Hole Binary in Braneworld Gravity}

In this section, we derive the rate of change of the orbital period of 
a binary system that harbors a black hole following closely the works of 
Will \& Zaglauer (1989) and 
Psaltis (2007b). In our treatment, we also include the effect of CFT 
emission of the black hole in the extra dimension.

For a black hole of mass $m_{1}$ with a companion star of mass $m_{2}$ on a 
circular 
orbit, the rate of change of the orbital angular momentum, 
$J\equiv\mu\sqrt{Gma}$, is\[
\frac{\dot{J}}{J}=\frac{1}{J}\frac{\partial J}{\partial m_{1}}\dot{m_{1}}+\frac{1}{J}\frac{\partial J}{\partial m_{2}}\dot{m_{2}}+\frac{1}{J}\frac{\partial J}{\partial a}\dot{a}\]
\begin{equation}
=\left(1-\frac{1}{2}\frac{m_{1}}{m_{1}+m_{2}}\right)\frac{\dot{m_{1}}}{m_{1}}+\left(1-\frac{1}{2}\frac{m_{2}}{m_{1}+m_{2}}\right)\frac{\dot{m_{2}}}{m_{2}}+\frac{1}{2}\frac{\dot{a}}{a},\label{jdotj}\end{equation}
where $m\equiv m_{1}+m_{2}$, $\mu\equiv m_{1}m_{2}/m$, and $a$ is the 
semi-major axis.
We set $m_{1}=qm_{2}$ and $\dot{m_{1}}=-\beta\dot{m_{2}}-\dot{M}$,
where $\dot{M}$ is the rate of black-hole evaporation into the
higher-dimensional bulk. We then obtain\begin{equation}
\frac{\dot{J}}{J}=\left(1-\frac{\beta}{q}-\frac{1}{2}\frac{1-\beta}{1+q}\right)\frac{\dot{m_{2}}}{m_{2}}-\left(1+\frac{1}{2}\frac{q}{1+q}\right)\frac{\dot{M}}{m_{1}}+\frac{1}{2}\frac{\dot{a}}{a}.\label{jdotjbq}\end{equation}

Angular momentum may be lost because of mass loss
from the system or because of the effect of magnetic braking. This leads to\begin{equation}
\frac{\dot{J}}{J}=j_{{\rm w}}(1-\beta)\frac{1+q}{q}\frac{\dot{m_{2}}}{m_{2}}+\frac{\dot{J}_{{\rm mb}}}{J},\label{jdotjjw}\end{equation}
where $j_{\rm{w}}$ is the specific angular momentum carried away by the
stellar wind in units of $2\pi a^{2}/P$, $P$ is the orbital
period and $\dot{J}_{{\rm mb}}/J$ is the rate of angular momentum loss
due to magnetic braking. Following Rappaport, Verbunt, \& Joss (1983) 
we estimate the corresponding torque by the empirical expression
\begin{equation}
\tau_{{\rm mb}}\equiv\dot{J}_{{\rm mb}}\simeq-3.8\times10^{-30}m_{2}R_\odot^{4}\left
(\frac{R_{2}}{R_\odot}\right)^{\gamma}\omega^{3}\, {\rm dyn\, cm}.\label{taumb}\end{equation}
Here, $\omega$ is the angular frequency of the secondary, $\gamma$
is a parameter that characterizes the strength of the magnetic braking,
and $R_{2}$ is the radius of the stellar Roche lobe which is assumed to be filled
at all times (Eggleton 1983),\begin{equation}
R_{2}=\frac{0.49q^{-2/3}}{0.6q^{-2/3}+\ln(1+q^{-1/3})}a.\label{rochelobe}\end{equation}
Using the expression for the orbital period\begin{equation}
\frac{P}{2\pi}=\frac{m}{m_{1}^{3}m_{2}^{3}}J^{3}G^{-2},\label{period}\end{equation} we can evaluate $\dot{J}_{{\rm mb}}/J$ as
\[\frac{\dot{J}_{{\rm mb}}}{J}=C\frac{G(m_{1}+m_{2})^{2}}{m_{1}}\left[\frac{0.49q^{-2/3}}{0.6q^{-2/3}+\ln(1+q^{-1/3})}\right]^{\gamma}\]
\begin{equation}
\times\left[\frac{\sqrt{G(m_{1}+m_{2})}}{2\pi}P\right]^{\frac{2}{3}(\gamma-5)},\label{jdotjmbperiod}\end{equation}
with\begin{equation}
C\equiv-3.8\times10^{-30}R_\odot^{4-\gamma}.\label{c}\end{equation}
Additionally, from the period equation (\ref{period}) together with our
 expressions for $m_{1}$ and $\dot{m_{1}}$ we find\[
\frac{\dot{P}}{P}=\frac{3}{2}\frac{\dot{a}}{a}-\frac{1}{2}\frac{-\beta\dot{m_{2}}-\dot{M}+\dot{m_{2}}}{m_{1}+m_{2}}\]
\begin{equation}
=-\frac{1}{2}\frac{1-\beta}{1+q}\frac{\dot{m_{2}}}{m_{2}}+\frac{1}{2}\frac{q}{1+q}\frac{\dot{M}}{m_{1}}+\frac{3}{2}\frac{\dot{a}}{a}.\label{pdotp}\end{equation}
Using\begin{equation}
\frac{\dot{q}}{q}=-\frac{\beta+q}{q}\frac{\dot{m_{2}}}{m_{2}}-\frac{\dot{M}}{m_{1}},\label{qdotq}\end{equation}
we obtain the rate of change of the radius of the companion\[
\frac{\dot{R_{2}}}{R_{2}}=\frac{\dot{a}}{a}+\frac{2}{3}\frac{\beta+q}{q}\left[1-\frac{0.6+0.5q^{1/3}(1+q^{-1/3})^{-1}}{0.6+q^{2/3}\ln\left(1+q^{-1/3}\right)}\right]\frac{\dot{m_{2}}}{m_{2}}\]
\begin{equation}
+\frac{2}{3}\left[1-\frac{0.6+0.5q^{1/3}(1+q^{-1/3})^{-1}}{0.6+q^{2/3}\ln\left(1+q^{-1/3}\right)}\right]\frac{\dot{M}}{m_{1}}.\label{rdotrq}\end{equation}

The third effect that dictates the change of the orbital period in a binary 
system is the evolution of the companion star. As the secondary leaves the
main sequence and starts to burn helium, it expands rapidly. Following 
Webbink, Rappaport, \& Savonije (1983) and Verbunt (1993) we estimate the 
rate of change of the radius of a star leaving the main sequence 
as\begin{equation}
\left(\frac{\dot{R}_{2}}{R_{2}}\right)_{{\rm ev}}=\left(c_{1}+2c_{2}y+3c_{3}y^{2}\right)\frac{\dot{M}_{{\rm c}}}{M_{{\rm c}}}. \label{rdotev}\end{equation}
Here, $M_{{\rm c}}$ is the core mass of the companion, $y\equiv\ln(M_{{\rm c}}/0.25M_\odot)$,
and $c_{1}$, $c_{2}$ and $c_{3}$ are constants that depend on the
composition of the core. The core mass changes in time according to
(Verbunt 1993)\begin{equation}
\dot{M}_{{\rm c}}\simeq1.37\times10^{-11}\left(\frac{L_{2}}{L_\odot}\right)M_\odot\, {\rm yr^{-1}}.\label{mcdot}\end{equation}
In this expression, $L_{2}$ is the luminosity of the companion, which is determined by the 
core mass (Webbink et al. 1983) according to the empirical 
relation\begin{equation}
\ln\left(\frac{L_{2}}{L_\odot}\right)=a_{0}+a_{1}y+a_{2}y^{2}+a_{3}y^{3},\label{lnl}\end{equation} with $a_{0}$, $a_{1}$, $a_{2}$, and $a_{3}$ constants depending
on the core composition. Combining equations (\ref{rdotev}) -- (\ref{lnl}) 
leads to
\[\left(\frac{\dot{R}_{2}}{R_{2}}\right)_{{\rm ev}}\simeq1.37\times10^{-11}\times4^{a_{1}}\left(c_{1}+2c_{2}y+3c_{3}y^{2}\right)\]
\begin{equation}e^{a_{0}+a_{2}y^{2}+a_{3}y^{3}}\left(\frac{M_{{\rm c}}}{M_\odot}\right)^{a_{1}-1}\, {\rm yr^{-1}}.\label{rdotrmc}\end{equation}

We now define the adiabatic index for the companion star as
\begin{equation} \xi_{{\rm ad}}\equiv\frac{d\ln R_{2}}{d\ln m_{2}}\label{xi}\end{equation}
and obtain\[
\frac{\dot{a}}{a}=\left[\xi_{{\rm ad}}-\frac{2}{3}\frac{\beta+q}{q}\left(1-\frac{0.6+0.5q^{1/3}(1+q^{-1/3})^{-1}}{0.6+q^{2/3}\ln\left(1+q^{-1/3}\right)}\right)\right]\frac{\dot{m_{2}}}{m_{2}}\]
\begin{equation}
-\frac{2}{3}\left[1-\frac{0.6+0.5q^{1/3}(1+q^{-1/3})^{-1}}{0.6+q^{2/3}\ln\left(1+q^{-1/3}\right)}\right]\frac{\dot{M}}{m_{1}}-\left(\frac{\dot{R}_{2}}{R_{2}}\right)_{\rm{ev}}.\label{adota}\end{equation}
In the following we will estimate the value of $\xi_{{\rm ad}}$ from the expressions for the stellar radius $R$ and mass $m_2$ given by Kalogera \& Webbink (1996).

Combining the equations we derived above, we obtain the rate of change of 
the orbital period of the binary\[
\frac{\dot{P}}{P}=Q_{0}\frac{\dot{M}}{m_{1}}+Q_{2}\frac{(m_{1}+m_{2})^{2}}{m_{1}}\left[\frac{0.49q^{-2/3}}{0.6q^{-2/3}+\ln(1+q^{-1/3})}\right]^{\gamma}\]
\[\times\left[\frac{\sqrt{G(m_{1}+m_{2})}}{2\pi}P\right]^{\frac{2}{3}(\gamma-5)}\]
\begin{equation}
+Q_{3}\left(c_{1}+2c_{2}y+3c_{3}y^{2}\right)e^{a_{0}+a_{2}y^{2}+a_{3}y^{3}}\left(\frac{M_{{\rm c}}}{M_{\odot}}\right)^{a_{1}-1}.\label{pdotpQ}\end{equation}
In this equation we have introduced the quantities
\[Q_{0}\equiv\frac{1}{2}\frac{1-\beta}{1+q}\frac{1+\frac{1}{2}\frac{q}{1+q}+\frac{1}{3}\mathcal{A}}{D}+\frac{1}{2}\frac{q}{1+q}\]
\begin{equation}
+\frac{3}{2}\frac{\left(\frac{2}{3}\frac{\beta+q}{q}\mathcal{A}-\xi_{{\rm ad}}\right)\left(1+\frac{1}{2}\frac{q}{1+q}+\frac{1}{3}\mathcal{A}\right)}{D}-\mathcal{A},\label{Qnaught}\end{equation}
\begin{equation}
Q_{2}\equiv\frac{C}{D}G\left(\frac{1}{2}\frac{1-\beta}{1+q}+\frac{\beta+q}{q}\mathcal{A}-\frac{3}{2}\xi_{{\rm ad}}\right),\label{Qtwo}\end{equation}
\[Q_{3}\equiv1.37\times10^{-11}\]
\begin{equation}
\times4^{a_{1}}\left[\frac{1}{4D}\frac{1-\beta}{1+q}+\frac{1}{2D}\left(\frac{\beta+q}{q}\mathcal{A}-\frac{3}{2}\xi_{{\rm ad}}\right)-\frac{3}{2}\right],\label{Qthree}\end{equation}
\begin{equation}\mathcal{A}\equiv1-\frac{0.6+0.5q^{1/3}(1+q^{-1/3})^{-1}}{0.6+q^{2/3}\ln\left(1+q^{-1/3}\right)},\label{A}\end{equation} and
\begin{equation}
D\equiv j_{{\rm w}}(1-\beta)\frac{1+q}{q}-1+\frac{\beta}{q}+\frac{1}{2}\frac{1-\beta}{1+q}-\frac{1}{2}\left(\xi_{{\rm ad}}-\frac{2}{3}\frac{\beta+q}{q}\mathcal{A}\right).\label{D}\end{equation}

For the ``evaporation'' of the black hole due to the emission of CFT modes, which are the dual description of the infinite dimension, we use (Emparan et al. 2003)\begin{equation}
\dot{M}=2.8\times10^{-3}\left(\frac{M_\odot}{m_{1}}\right)^{2}\left(\frac{L}{1~{\rm mm}}\right)^{2}M_\odot\, {\rm yr^{-1}},\label{Mdot}\end{equation}
where $L$ is the asymptotic AdS radius of curvature. Defining\begin{equation}
Q_{1}\equiv2.8\times10^{-3}Q_{0}\, {\rm yr^{-1}},\label{Qone}\end{equation}
we arrive at our final equation for the orbital period evolution:\[
\frac{\dot{P}}{P}=Q_{1}\left(\frac{M_\odot}{m_{1}}\right)^{3}\left(\frac{L}{1~{\rm mm}}\right)^{2}\]
\[
+Q_{2}\frac{(m_{1}+m_{2})^{2}}{m_{1}}\left[\frac{0.49q^{-2/3}}{0.6q^{-2/3}+\ln(1+q^{-1/3})}\right]^{\gamma}\]
\[\times\left[\frac{\sqrt{G(m_{1}+m_{2})}}{2\pi}P\right]^{\frac{2}{3}(\gamma-5)}\]
\begin{equation}
+Q_{3}\left(c_{1}+2c_{2}y+3c_{3}y^{2}\right)e^{a_{0}+a_{2}y^{2}+a_{3}y^{3}}\left(\frac{M_{{\rm c}}}{M_\odot}\right)^{a_{1}-1}.\label{final}\end{equation}

For our analysis it is important that the companion star remains in contact with its Roche lobe. Should the black-hole evaporation be so strong that magnetic braking is negligible (as well as stellar evolution), then the Roche lobe of the secondary will grow and the system will eventually get out of contact. This, however, can only occur on timescales of at least $10^8~{\rm yr}$ which is much larger than the observational timescales of interest. Studying other implications of the loss of contact for the evolution of the black-hole binary is beyond the scope of this paper.

\section{Results}

In this section, we investigate the potential of the currently known black-hole binary systems to constrain the rate of black-hole evaporation into higher dimensions. We will start with a general discussion of the various systems and then focus on the system A0620$-$00 in particular.

First we investigate which term in equation (\ref{final}) dominates the period evolution for a given period $P$, black hole and companion masses $m_1$,  $m_2$, respectively, and curvature radius $L$. Figure~1 shows the orbital periods versus the companion masses of the observed systems. On the same graph we plot the curves along which the evaporation term equals the magnetic braking term for $L=0.1~{\rm \mu m}$, $L=1~{\rm \mu m}$, $L=10~{\rm \mu m}$, and $L=100~{\rm \mu m}$. The evaporation dominates above the lines, whereas below the lines the magnetic braking dominates. Recent numerical simulations (Yungelson \& Lasota 2008) showed that the expression for magnetic braking is actually overestimated for low-mass black-hole binaries, which further increases the predominance of the evaporation term. The parameters we used in this figure are $\xi_{{\rm ad}}=0.8$, $\beta=0$, $j_{\rm w}=0$, and $\gamma=0$, and we set the black-hole mass to a nominal value of 10$M_{\odot}$.

For companion masses $\gtrsim 1M_{\odot}$, there exists a maximum period beyond which systems contain companion stars that have evolved past the base of the giant branch (see the curve marked BGB in Figure~1). For these systems, the evolution of the companion star completely dominates the rate of change of the orbital period for any plausible value of the asymptotic curvature radius $L$.

We find that two sources have relatively large orbital periods and are probably evolved, while the other systems group around the line that corresponds to $L=1~{\rm \mu m}$ and between the lines corresponding to $L=10~{\rm \mu m}$ and $L=100~{\rm \mu m}$. Among the latter we identify the system A0620$-$00 as a promising candidate both because of the theoretical expectation shown in this figure and because of the extensive historical monitoring of its orbital period. We analyze this source more closely in the following.

\begin{figure}
\plotone{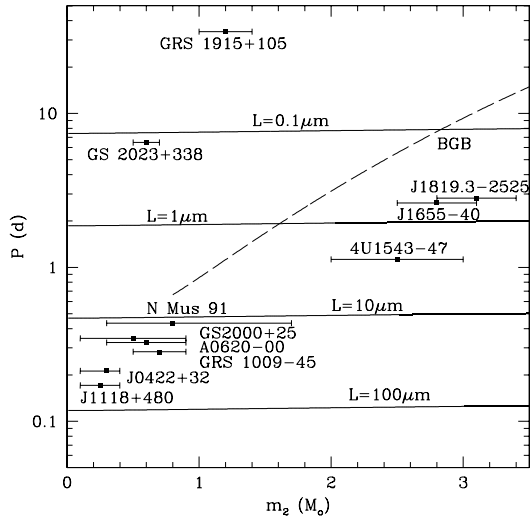}
\caption{The orbital period $P$ of observed black-hole binary systems versus the mass $m_2$ of the companion star. Four separatrices are shown for different values of the asymptotic AdS curvature radius $L$ for a nominal black-hole mass of 10$M_{\odot}$. Below the lines, magnetic braking dominates. Above the lines, black-hole evaporation dominates. Binaries above the curve marked BGB contain companions beyond the base of the giant branch; for these systems, the evolution of the companion completely dominates the orbital evolution of the binary. The parameters for this graph are $\xi_{{\rm ad}}=0.8$, $\beta=0$, $j_{\rm w}=0$, and $\gamma=0$. Note that the separatrices only depend weakly on $m_2$ as long as $\gamma=0$. \label{fig1}}
\end{figure}

\begin{table}[t]
\centerline{
\footnotesize
\begin{tabular}{lcccl}
\multicolumn{5}{c}{Table 1: Observed Properties of X-Ray 
Binaries\tablenotemark{a}}\\
\hline \hline
X-Ray Binary & $P\ (h)$ & $q$ & $m_1\ (M_{\odot})$\\
\hline
GRS1915+105 &816 &12 &14$\pm$4\\
J1118+480 &4.1 &$\sim$20 &6.8$\pm$0.4\\
GS2023+338 &155.3 &17$\pm$1 &12$\pm$2\\
GS2000+25 &8.3 &24$\pm$10 &10$\pm$4\\
H1705-25 &12.5 &$>$19 &6$\pm$2\\
GRS1009-45 &6.8 &7$\pm$1 &5.2$\pm$0.6\\
N Mus 91 &10.4 &6.8$\pm$2 &$6^{+5}_{-2}$\\
A0620-00 &7.8 &17$\pm$1\tablenotemark{b} &10$\pm$5\\
J0422+32 &5.1 &$9.0^{+2.2}_{-2.7}$ &4$\pm$1\\
J1819.3-2525 &67.6 &2.31$\pm$0.08 &7.1$\pm$0.3\\
J1655-40 &62.9 &2.39$\pm$0.15 &6.6$\pm$0.5\\
4U1543-47 &27.0 &3.6$\pm$0.4 &9.4$\pm$1\\
\hline
\end{tabular}}
\tablenotetext{a} {Most data compiled by Charles \& Coe (2006)}
\tablenotetext{b} {Neilsen, Steeghs \& Vrtilek (2008)}
\end{table}

The short-period (0.32 d) binary A0620$-$00 has a secondary that
resembles a main sequence star, as indicated by its spectrum and by its
chemical abundances (Gonz\'{a}lez Hern\'{a}ndez et al. 2004).  Its mass
and radius are also comparable to that of a main-sequence K4 star (Marsh
et al. 1994).  In particular, the mean density of this
Roche-lobe-filling secondary, which is precisely determined by its
orbital period (Frank et al. 2002), is only $\sim 25$\% below that of a
normal K4 dwarf.  However, the secondary of A0620$-$00 is not a normal
star, given the extraordinary evolutionary history of this black-hole
binary system (e.g., de Kool et al. 1986).  Nevertheless, for our
purposes the secondary functions like a main-sequence star: it is not
evolving on a nuclear time scale and the system is kept in contact by
magnetic braking (Justham et al. 2006).

Thus we can neglect the evolution term in equation (\ref{final}) and plot the expected rate of change of its orbital period $P$ as a function of the asymptotic AdS curvature radius $L$ (see Figure~2). The parameters for this plot are $\beta=0$, $j_{\rm w}=0$, $\gamma=0$, and $\xi_{{\rm ad}}=0.8$. We see that for $L\lesssim20~\mu{\rm m}$ the magnetic braking dominates and the rate of orbital period change is constant because it is independent of $L$. For $L\gtrsim20~\mu{\rm m}$ the black-hole evaporation dominates and the orbital period derivative increases with increasing AdS curvature as expected. This shows that A0620$-$00 theoretically allows for a constraint on $L$ as low as $20~\mu{\rm m}$, assuming that $m_1=10M_{\odot}$. Since $m_1$ has only been measured to an accuracy of $\pm$50\%, the constraint can even be reduced to a few microns. We will return to the question of the black-hole mass in the next section.

\begin{figure}
\plotone{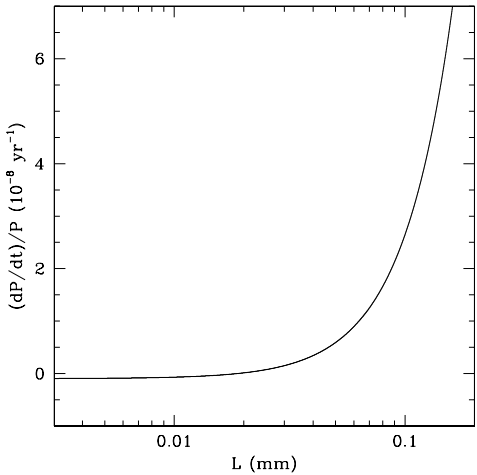}
\caption{The rate of change of the orbital period $P$ of the binary system A0620$-$00 versus the asymptotic curvature radius $L$ in the extra dimension. The parameters are $\xi_{{\rm ad}}=0.8$, $\beta=0$, $j_{\rm w}=0$, and $\gamma=0$. The transition from predominant magnetic braking (constant negative rate) to predominant black-hole evaporation (positive and rapidly increasing rate) occurs at $L\simeq20~\mu{\rm m}$.\label{fig2}}
\end{figure}

In order to determine the dependence of the orbital period evolution on the parameters $j_{\rm w}$, $\beta$, and $\gamma$, we plot the rate of change of the orbital period as a function of one parameter while holding the others constant. In all plots, we set $\xi_{{\rm ad}}=0.8$ (estimated from Kalogera \& Webbink 1996) for the companion mass in this system and evaluate the period evolution rate at the current experimental upper limit of the AdS curvature of $L=44~\mu{\rm m}$ (Kapner et al. 2007). Figure~3 shows the dependence of the rate of change of the orbital period on the parameters $j_{\rm w}$, $\beta$, and $\gamma$, respectively. First we note that for large values of the parameters the period increases. This behavior is entirely due to the high mass ratio $q=m_1$/$m_2$ measured for this source; for other sources with substantially smaller mass ratios, the behavior is not monotonic. Furthermore, we find that the rate of change of the orbital period is the smallest when $j_{\rm w}=0$ (no angular momentum loss due to stellar wind), $\beta=0$ (no accretion), and $\gamma=0$. We choose these values for the respective parameters in the following discussion, where we are aiming to calculate a lower limit on the expected rate of change of the orbital period.

\begin{center}
\begin{figure}
\psfig{figure=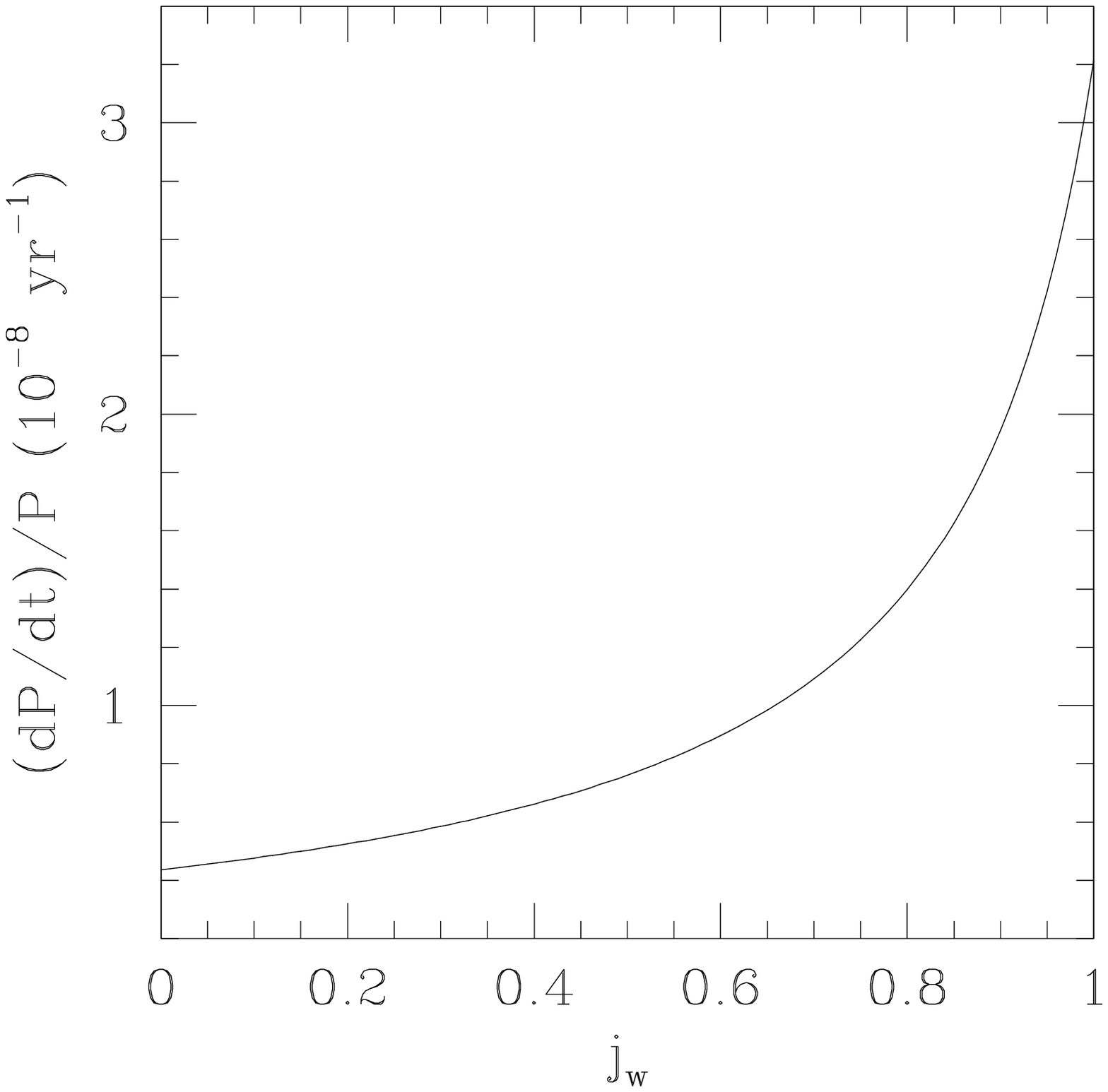,height=2.1in}
\psfig{figure=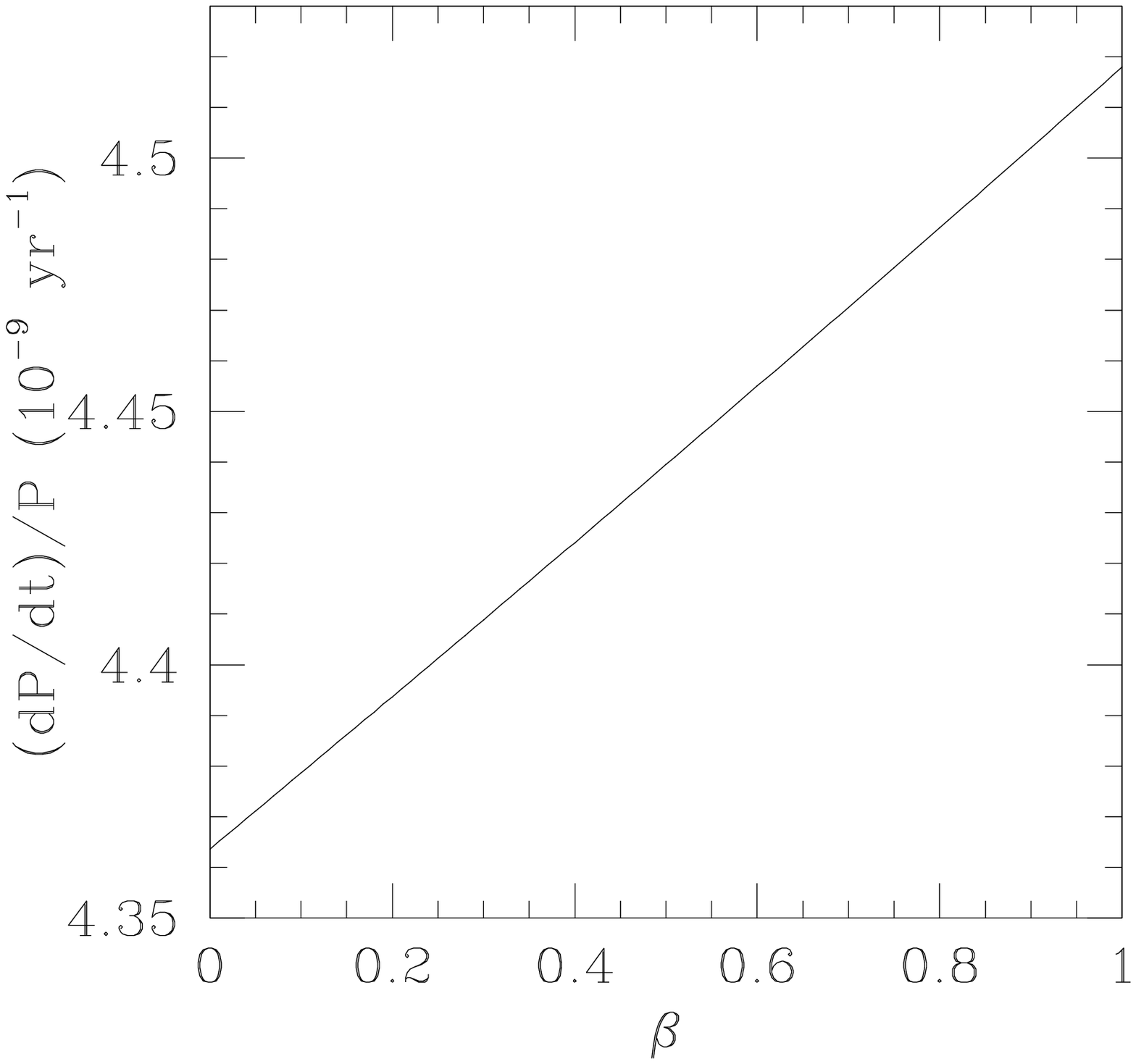,height=2.1in}
\psfig{figure=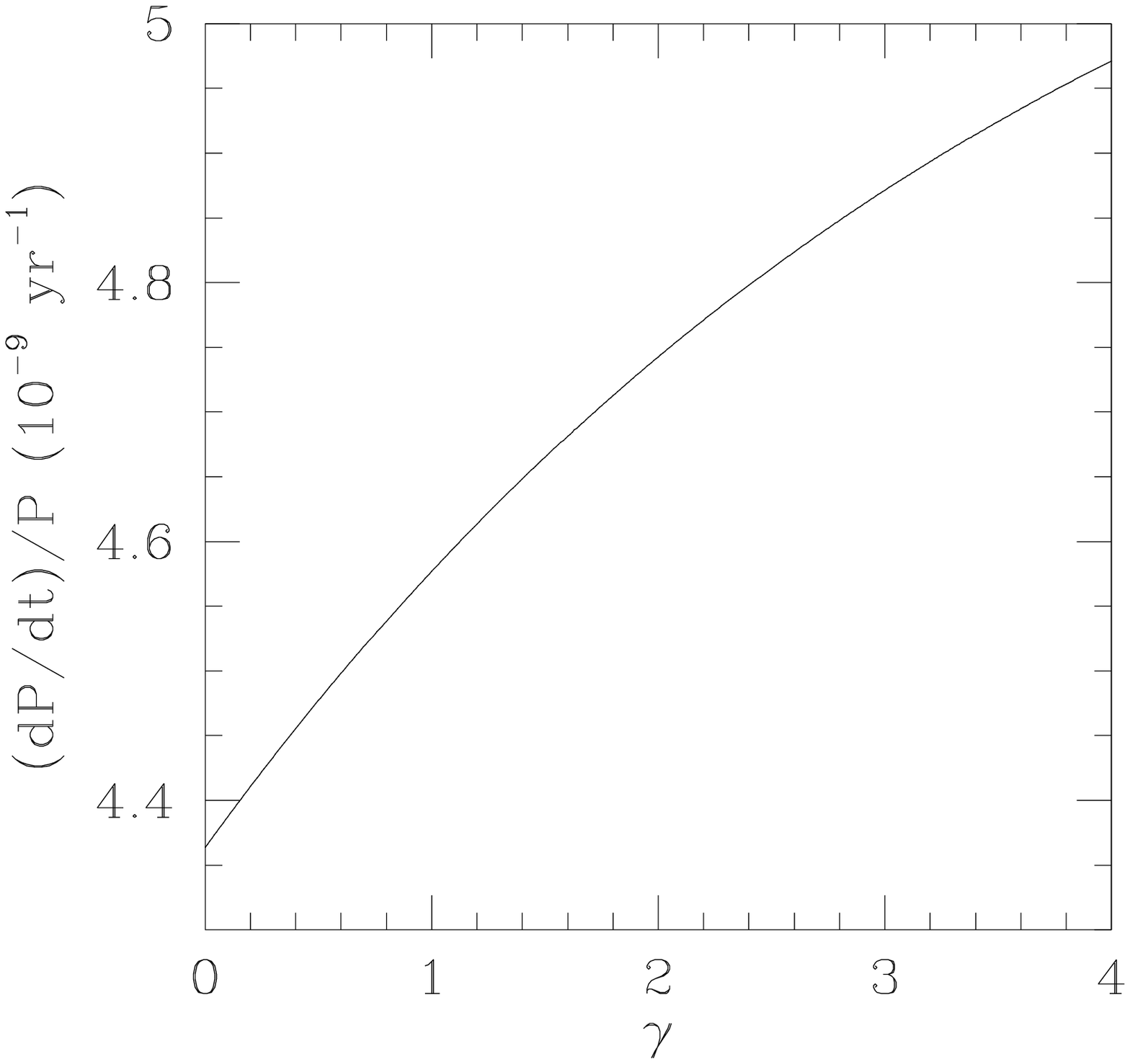,height=2.1in}
\caption{The rate of change of the orbital period $P$ (in years) of the binary system A0620$-$00 versus the specific angular momentum removed by the wind $j_{\rm w}$, the accretion parameter $\beta$, and the magnetic braking parameter $\gamma$, for $L=44~\mu{\rm m}$ and $\xi_{\rm ad}=0.8$. On varying one parameter, the others are held constant at the respective values $j_{\rm w}=0$, $\beta=0$, and $\gamma=0$.}
\end{figure}
\end{center}

\section{Data}

In this section we use previously published measurements of the orbital period of
A0620$-$00 to set an upper limit on the size $L$ of the asymptotic
curvature in the extra dimension. First we determine that limit using
the best fit black-hole mass of $m_1=10M_{\odot}$, and then we proceed with
an analysis of $L$ for different black-hole masses.

The orbital period of A0620$-$00 ($P = 0.32$~d) has been measured several
times during the past two decades.  A convenient orbital phase reference
is the time of maximum radial velocity $T_{\rm 0}$.  Four measured
values of $T_{\rm 0}$, which span 22 years, are given in
Table 2.  These times and the individual determinations of the orbital
period $P$ uniquely determine the cycle number $n$. The table also
gives the calculated times of maximum velocity based on a simple
ephemeris with a constant orbital period and referenced to the most
precise and recent determination of $T_{\rm0}$ (see footnote $a$ of Table
2).  The differences between the observed and calculated times
are smaller than their corresponding uncertainties, and thus there is
no evidence for any change in the orbital period during the past 22
years.  Note that with modern telescopes and instrumentation one can
routinely achieve a precision of several seconds in $T_{\rm 0}$ (see the
last entry in the table) in $\sim 10$~hours of radial velocity
observations of a source like A0620$-$00 (Neilsen et al. 2008).

Of interest to us is a secure limit on the rate of change of the orbital
period. A constant rate of change of the period will result in a quadratic
variation in $T_{\rm0}$ (e.g., Kelley et al. 1983).  The time of the $n$th 
value of $T_{\rm 0}$ is then given by

\begin{equation}
t_{\rm n} = t_{\rm 0} + Pn+\frac{1}{2}P\dot{P}n^2,\label{Pdotexp}
\end{equation}

\noindent where $P$ and $\dot{P}$ are the orbital period and its
derivative, respectively, at time $t_{\rm 0}$; $n$ is the orbital cycle
number.  Following the standard procedure and using the IDL routine {\it
curvefit}, we fitted for the three parameters $t_{\rm 0}$, $P$ and
$\dot{P}$ using the four observed values of $T_{\rm 0}$ given in Table
2 (Figure~4).  The fit yields $\dot{P}=(-1.66\pm2.64)\times10^{-11}~{\rm s/s}$.
Thus, using a 3$\sigma$-upper limit, the period
derivative is constrained within the interval $-9.58\times10^{-11}~{\rm s/s}<\dot{P}<6.26\times10^{-11}~{\rm s/s}$. Next we plot the
rate of change of the orbital period versus the AdS curvature for the
values of the parameters $\beta$, $j_{\rm w}$, and $\gamma$ that lead to the
lowest limit of the rate of change of the orbital period for a given value of the asymptotic curvature radius $L$ (Figure~5). For $L\gtrsim20~{\rm \mu m}$, the rate of change of the orbital period is smallest for the set of parameters $j_{\rm w}=0$, $\beta=0$, and $\gamma=0$, while for $L\lesssim20~{\rm \mu m}$, the parameters $j_{\rm w}=1$, $\beta=0$, and $\gamma=0$ minimize the orbital period evolution. In a narrow intermediate region, the corresponding set of parameters is $\beta=1$, and $\gamma=0$, while $j_{\rm w}$ is arbitrary. Since the
measured upper limit on the period derivative in equation
(\ref{Pdotexp}) marks the largest time change of the orbital period, the
intersection point of this line with the graph of the smallest rate of change of the orbital period  places an upper limit on the asymptotic AdS curvature radius of $L\leq161~\mu{\rm m}$ (Figure~5).

\begin{figure}
\plotone{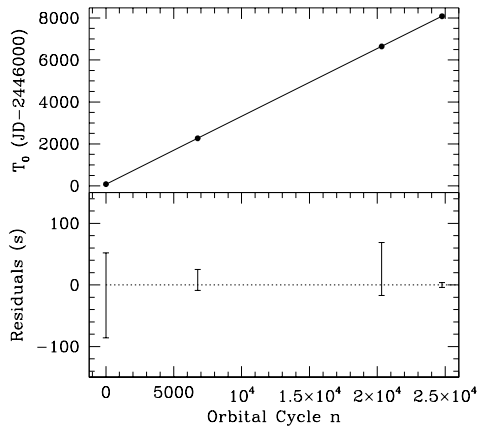}
\caption{The time of maximum radial velocity $T_0$ and residuals versus the orbital cycle number $n$ for the binary A0620$-$00.}
\end{figure}

Since the upper limit on the asymptotic curvature radius depends strongly on the mass of the black hole, which has only been measured to an accuracy of $
\pm50\%$, we plot in Figure~6 the upper limit $L_{{\rm max}}$ versus the black-hole
mass for different limits on the orbital period change. For the upper
curve we used the current limit on the orbital period evolution from
the fit using equation (\ref{Pdotexp}), whereas for the lower curve we used a 
value that is a factor of 10 lower. We see that even
the current table-top limit of the asymptotic AdS curvature radius 
$L=44~\mu{\rm m}$ can be improved if the upper limit of
the rate of change of the orbital period can be reduced to 10\% of the
current value and if the black-hole mass is smaller than 9$M_{\odot}$.
In Figure~7 we plot the smallest rate of change of the orbital period for black-hole masses (from left to right) $m_1=5,~10,~{\rm and}~15M_{\odot}$, respectively. The parameters for each curve are chosen as in Figure~5.

There are ample opportunities to substantially improve the above limit
on the asymptotic curvature radius $L$ using our method.  For example, because the uncertainty in
the orbital period derivative depends quadratically on the orbital cycle number $n$, even a single future
observation of the binary A0620$-$00 that extends the 22-year baseline by just 5
years (i.e., $n = 30423$) with a precision of 4 s would reduce the error
in the orbital period derivative $\dot{P}$ by a factor of six.  Furthermore, independent limits of
comparable quality on $\dot{P}$ and $L$ could be obtained by monitoring
the ephemerides of several other black-hole X-ray binary systems (e.g., GRS
1124--683, XTE J1118+480, and 4U 1543--47; Remillard \& McClintock 2006).

\begin{table*}[t]
\begin{center}
\begin{tabular}{ccccc}
\multicolumn{5}{c}{Table 2: Observed and Computed Heliocentric Times of
  Maximum Velocity for A0620--00} \\
\hline \hline
\multicolumn{1}{c}{Orbital Cycle}& 
\multicolumn{1}{c}{$T_0$ in JD}&
\multicolumn{1}{c}{$T_{0}(n)$ in JD\tablenotemark{a}}&
\multicolumn{1}{c}{$T_0-T_{0}(n)$}&
\multicolumn{1}{c}{Reference\tablenotemark{b}} \\
\multicolumn{1}{c}{$n$}&
\multicolumn{1}{c}{observed}&
\multicolumn{1}{c}{computed}&
\multicolumn{1}{c}{(s)}&
\multicolumn{1}{c}{} \\

\hline

0&      2,446,082.7481$\pm$0.0008&                   2,446,082.7483&   $-17\pm69$&  1 \\

6764&   2,448,267.6155$\pm$0.0002\tablenotemark{c}&  2,448,267.6154&   $8\pm17$&   2 \\

20321&  2,452,646.7173$\pm$0.0005&                   2,452,646.7170&   $26\pm43$&  3 \\

24773&  2,454,084.77560$\pm$0.00005&                 2,454,084.77560&  $0\pm4$&    4 \\     

\hline 
\end{tabular} 
\tablenotetext{a}{$T_{0}(n)$ = JD 2,454,084.77560 -- (24773 -- $n$)$\times$0.32301406.}
\tablenotetext{b}{REFERENCES: (1) McClintock \& Remillard 1986; (2) Orosz et al.\ 1994; (3) Shahbaz et al.\ 2004; (4) Neilsen et al.\ 2007.}
\tablenotetext{c}{$T_0$ corrected to date of observation on 1991 January 11 using ephemeris in reference (1).}
\end{center}
\end{table*}

\begin{figure}
\plotone{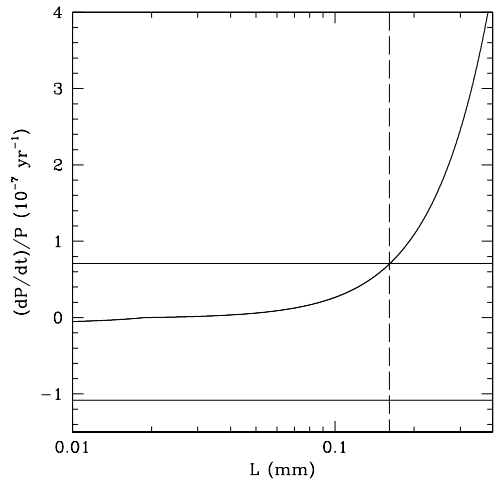}
\caption{The rate of change of the orbital period $P$ of the binary system A0620$-$00 as a function of the asymptotic AdS curvature radius $L$ for a black-hole mass of 10$M_{\odot}$. For positive values of the period derivative, this line represents the lower limit among all possible values of the period evolution. The 3$\sigma$-error bars of the observed orbital period derivative are shown as horizontal lines. The intersection point of the upper limit on $\dot{P}/P$ with the lower limit curve marks our constraint on the asymptotic curvature in the bulk of $L=161~\mu{\rm m}$ (vertical line).\label{fig6}}
\end{figure}

\begin{figure}
\plotone{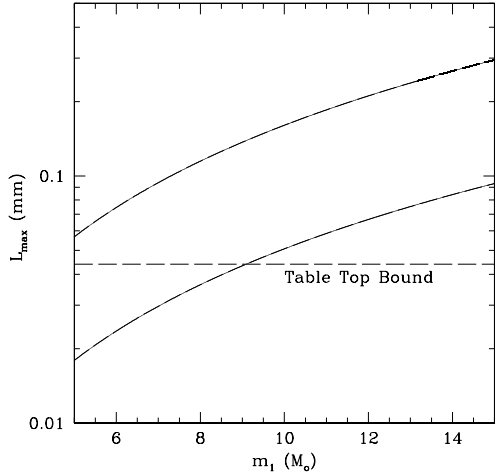}
\caption{The upper limit on the asymptotic AdS curvature radius $L$ as a function of the black hole mass $m_1$ of A0620$-$00 using the current lower limit on the orbital period evolution (upper curve) and on 10\% of that limit (lower curve). The dashed line shows the current upper limit on $L$ from table-top experiments.\label{fig7}}
\end{figure}

\section{Discussion}

For a binary system consisting of a black hole and a companion star we derived the rate of change of the orbital period in the RS2 braneworld gravity model incorporating the emission of CFT modes in the extra dimension. Magnetic braking and the evolution of the companion star can also change the orbital period, but they are negligible if the secondary is a main-sequence star and if the asymptotic AdS curvature radius $L$ is large enough so that the evaporation dominates. Measuring the rate of change of the orbital period then allows us to constrain the asymptotic AdS curvature radius.

\begin{center}
\begin{figure}
\plotone{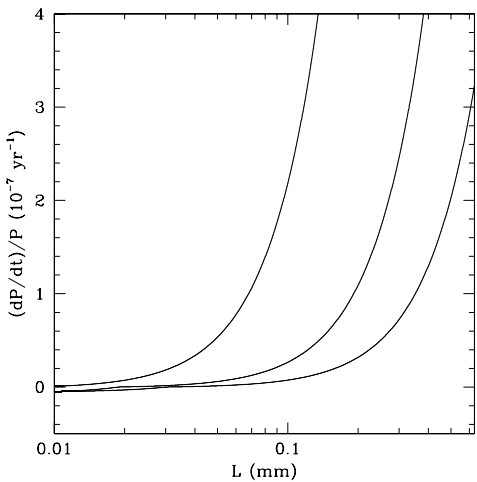}
\caption{The minimum rate of change of the orbital period of the binary A0620$-$00 versus the asymptotic AdS curvature radius $L$ for (from left to right) $m_1=5,~10,~15M_{\odot}$, respectively.}
\end{figure}
\end{center}

We analyzed in detail the binary system A0620$-$00 which is a good candidate for such a constraint for both theoretical and observational reasons. The evaporation term dominates the change of the orbital period as long as the asymptotic curvature radius is at least $\sim$20 microns large. Measurements of the orbital period over the last 20 years allow for a constraint of $L<161~\mu{\rm m}$ assuming a black-hole mass of 10$M_{\odot}$. Refining the measurement of the mass of the black hole and of the rate of change of the orbital period can further improve the constraint on the asymptotic curvature radius. As an example we showed that improving the measurement of the rate of change of the orbital period by one order of magnitude will constrain the AdS curvature radius to a value smaller that the current experimental limit $L=44~\mu{\rm m}$ (Kapner et al. 2007), provided the black-hole mass is measured not to exceed 9$M_{\odot}$.

\begin{center}
\begin{figure}
\plotone{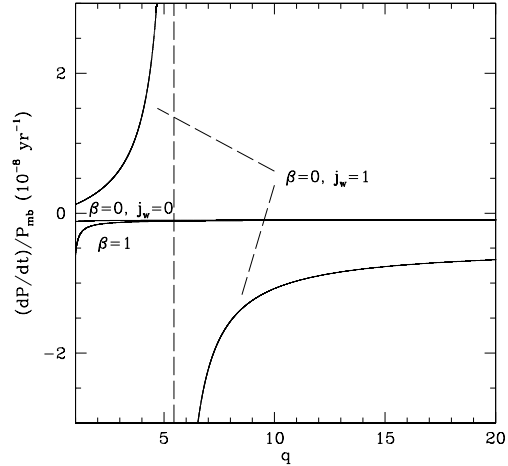}
\caption{The rate of change of the orbital period of the binary A0620$-$00 due to magnetic braking versus the mass ratio $q=m_1/m_2$ for several combinations of the parameters $\beta$ and $j_{\rm w}$. We set $m_1=10M_{\odot}$, $P=7.8~{\rm hr}$, $\xi_{\rm ad}=0.8$, and $\gamma=0$. For large values of the mass ratio, varying the parameters only changes the magnitude of the magnetic braking term but not its overall sign.}
\end{figure}
\end{center}

Considering the other known black-hole X-ray binaries, we see from Figure~1 that there are more systems which we can use in constraining the asymptotic curvature radius in the extra dimension. The requirement of unevolved secondaries rules out some of them, but several sources have the potential of a constraint on the curvature radius down to a few microns.

A very exciting aspect of our method is that it not only allows us to constrain the AdS radius $L$, but also to potentially measure it --- thereby giving evidence for new physics beyond general relativity and the standard model --- provided that the rate of change of the orbital period is measured to be positive. This is due to the fact that, for a black-hole binary with a sufficiently high mass ratio ($q=m_1/m_2$), magnetic braking can only shorten the orbital period. The sign of the magnetic braking term in equation (\ref{final}) and hence whether this effect leads to a positive and a negative change of the orbital period is determined exclusively by the prefactor $Q_2$. This factor depends only on the mass ratio $q$ and on the parameters $\beta$ and $j_{\rm w}$ assuming a fixed adiabatic index. In Figure~8 we plot the magnetic braking term for various combinations of the parameters $\beta$ and $j_{\rm w}$, and we set the black-hole mass and the orbital period to the nominal values $m_1=10M_{\odot}$ and $P=7.8~{\rm hr}$, respectively, as well as $\gamma=0$. The mass of the primary, the orbital period and the parameter $\gamma$ only effect the magnitude of the magnetic braking term but not its sign. For all curves we set $\xi_{\rm ad}=0.8$. We see that the magnetic braking term is negative for $q>5.5$ for any value of the parameters $\beta$, $j_{\rm w}$, and $\gamma$.

Thus we conclude that any measurement of a positive orbital period derivative in a black-hole X-ray binary with a mass ratio $q>5.5$ (assuming an unevolved companion star) is strong evidence for new gravitational physics and directly measures the asymptotic curvature radius $L$. Since our method is model-dependent (RS2), such a measurement, together with the table-top experiments, would even allow for a distinction between the ADD and the RS2 scenario. Current table-top experiments, that probe Newton's inverse square law in the sub-mm range, are insensitive to the underlying model, so that their results together with a measurement as discussed above would indeed allow for a distinction between ADD and RS2.

We would like to thank Nemanja Kaloper and Keith Dienes for fruitful discussions as well as Jack Steiner for his help with the IDL {\it curvefit} calculations. We also thank Saul Rappaport for his input on issues of black-hole binary evolution. This work was supported in part by the NSF CAREER award NSF 0746549.

\end{document}